# Toward Trusted Sharing of Network Packet Traces Using Anonymization: Single-Field Privacy/Analysis Tradeoffs

William Yurcik, Clay Woolam, Greg Hellings, Latifur Khan, Bhavani Thuraisingham
University of Texas at Dallas
<byurcik@gmail.com>  {cpw021000,gsh062000,lkhan,bhavani.thuraisingham}@utdallas.edu

## ABSTRACT
Network data needs to be shared for distributed security analysis. Anonymization of network data for sharing sets up a fundamental tradeoff between privacy protection versus security analysis capability. This privacy/analysis tradeoff has been acknowledged by many researchers but this is the first paper to provide empirical measurements to characterize the privacy/analysis tradeoff for an enterprise dataset. Specifically we perform anonymization options on single-fields within network packet traces and then make measurements using intrusion detection system alarms as a proxy for security analysis capability. Our results show: (1) two fields have a zero sum tradeoff (more privacy lessens security analysis and vice versa) and (2) eight fields have a more complex tradeoff (that is not zero sum) in which both privacy and analysis can both be simultaneously accomplished.

## Categories and Subject Descriptors
C.2.0 [**Computer-Communication Networks**]: General – *security and protection;* C.2.3 [**Computer-Communication Networks**]: Network Operations – *network monitoring;* C.2.m [**Computer-Communication Networks**]: Miscellaneous – *packet trace processing;* D.3.4 [**Programming Languages**]: Processors – *network trace rewriting;* K.6.5 [**Management of Computing and Information Systems**]: Security and Protection.

## 1. INTRODUCTION
Security operations staff use data, such as network packet traces, to help defend their own organizational networks. Since attackers typically attack across network boundaries and frequently change targets to attack within different security domains, effective protection requires defenders to look beyond their own organizational perimeter toward cooperation and data sharing with other organizations in order to defeat attackers. However, to date little or no data sharing has occurred between organizations due to practical concerns. Unfortunately, this is not also true for attackers who are quite efficient at sharing vulnerability and exploit information amongst themselves.[3]

The practical concerns which have prevented organizations from sharing network packet traces include the resources needed to prepare data for sharing and a valid fear that private and/or sensitive information in shared data may be misused to cause harm.[9] Examples of private/sensitive information in network packet traces that we may not want to reveal include personally-identifiable user information, user activities, system configurations, network topologies, network services, organizational defenses, and attack impacts – all valuable information to attackers.

We can prevent attackers from exploiting private/sensitive data within shared data by deleting it but the equally important challenge is for shared data to remain useful for collaborative security analysis – the reason for data sharing in the first place! Such collaborative security analysis may include incident detection, trend analysis, attack detection, black listing specific attackers and attacker techniques, and distinguishing/modeling normal versus suspicious network traffic patterns.

*How do we accomplish collaborative security analysis with shared data if organizations do not trust each other?* Absent an established trusted relationship, it is difficult for an organization to determine if a sharing party is a legitimate peer organization or an attacker. If trusted relationships can be built using available means then sharing data may not be worrisome. However, it may be the case that a significant number of the organizations seeking to share have not previously interacted. Internet security is very dynamic requiring many relationships such that building trusted relationships to all relevant organizations in near-real-time may not be feasible. Automation is needed.

Access control automation to shared data is possible for small groups but as the group size grows the feasibility of controlling access to shared data using access control mechanisms diminishes. All it would take is one organization in a group to have its access control credentials compromised or one organization in a group to turn traitor and then shared data would be exposed. Anonymization is an automated solution that can provide practical levels of assurance that shared data cannot be used to cause harm. However, the use of anonymization entails tradeoffs that must be tailored to the participating organizations and specific situations.

The fundamental tradeoff in data sharing between organizations using anonymization is the risk of valuable network data being unknowingly disclosed (privacy protection not stringent enough) versus valuable network data being needlessly deleted (security analysis maligned with privacy protection too stringent) – we refer to this as the privacy/analysis tradeoff. Since different organizations have security policies with different privacy requirements, multi-level anonymization options for each field provides flexibility in selecting anonymization schemes to more closely match real sharing environments between parties. While one possible approach is to protect privacy of data from all organizations at the highest possible level, the tradeoff between privacy protection and security analysis is such that the most stringent privacy protection anonymization options may significantly degrade (or even eliminate) possible insights from desired security analysis. There is no one-size-fits-all anonymization scheme that will work, so it is critical to have multi-level anonymization options to enable the tailoring of anonymization to different sharing situations.

In this paper we specifically focus on characterizing the privacy/analysis tradeoffs of anonymization options on single fields within a packet, with each field being independently acted upon by an anonymization tool. Characterizing these privacy/analysis tradeoffs is a first step toward facilitating trusted sharing of packet traces using anonymization. Future work will examine the privacy/analysis tradeoffs of multi-field packet trace anonymization – the amount of analysis required for the combinatorics of multi-field cases is beyond the scope of this paper. We feel it is prerequisite that the single-field case be analyzed and documented first (in this paper) before attempting to analyze the more complex multi-field case.

The remainder of this paper is organized as follows: Section 2 presents a survey of related work. Section 3 presents our experimental design including an overview of the experimental protocol, packet trace anonymization tool, the source dataset, and the security analysis system for generating metrics. Section 4 reports experimental research results. We conclude with a summary, conclusions, and future work in Section 5.

## 2. RELATED WORK

Trust management, as introduced by Blaze et al. [2] in 1996, is a unified approach to specifying/interpreting security policies, credentials, and relationships that enable security-critical events.[1] Trust management entails collecting information necessary to establish a trust relationship, dynamically monitor a trust relationship, and modify a trust relationship.[11] Various models for trust establishment have been proposed including: (1) public-key cryptography[7], (2) the resurrecting duckling model[26], and (3) the distributed trust model[1]. Both public-key cryptography and the resurrecting duckling model are traditional security mechanisms that assume trust can be built via out-of-band mechanisms or based on a priori knowledge.[22] However, these traditional security mechanisms fail when out-of-band mechanisms and a priori knowledge is not possible. For example, there is unlikely to be advanced agreement on a trusted third party between organizations in different domains that have never interacted before.

The distributed trust model for trust establishment can be further subdivided into distinct categories:[11]

1. evidence-based models in which entities establish trust relationships based on evidence such as keys

---

[1] Prior to 1996 some security solutions for distributed networked applications already had an implicit notion of trust management based on PGP or X.509 public key certificates.

2. recommendation-based models in which recommendations from intermediaries set up trust relationships between strangers
3. pseudonym-based models in which information obfuscated by anonymization set up trust relationships between cooperating entities such as entities involved in a transaction

This paper is seeking an automated way to build trust between distributed organizations so they can share network data based on the distributed trust model, particularly the pseudonym-based model. The Marsh thesis on computational artificial trust was the first work to seek solutions for building trust on the Internet without out-of-band mechanisms or a priori knowledge.[14] Computational artificial trust may instead be built via evidence from Internet interactions. However, it was found that the level of computational artificial trust is a tradeoff with other properties, most notably privacy.

In [24] Seigneur and Jensen present trust/privacy tradeoffs similar to the problem posed in this paper – the more knowledge of an entity, the more accurate the trustworthy assessment versus the less privacy for the entity (and vice versa). They propose the use of pseudonyms as a level of indirection which allows the formation of trust without exposing real-world identity. A protocol is presented for incrementally disclosing evidence linked to pseudonyms in an attempt to satisfy anonymity ("nymity") and utility thresholds. In [23] the idea of "trust transfer" is proposed in which trust is transferred via recommendations from a recommender to the subject with a trust engine developed for an anti-spam Email example.

[18] presents web service personalization versus privacy tradeoffs. While users desire personalized website services, to accomplish personalization invades privacy with user profiles created by monitoring, analyzing, and storing user data and activities. The authors propose a "mask" pseudonym that protects privacy by hiding the identity of individual users by classifying them into groups.

[5] introduces the *ACORN* distributed multi-agent architecture for managing information across networks. Trust in *ACORN* is built by implementing an anonymous service provider to protect privacy by disassociating agents and users.[5] A server anonymizes information agents before they are sent to the network and re-instantiates them on their return from the network.[6]

Initial attempts at privacy-preserving sharing of network data were deletion of private/sensitive data which in many cases also needlessly destroyed desired information and/or altered the data format of the rest of the log. Anonymization was identified simultaneously by many researchers as a better way to enable sharing of obfuscated network data that remains useful from an analytical point of view while still guaranteeing that private/sensitive information cannot be derived. For a more comprehensive treatment of network data anonymization for secure sharing see [25]. [31] presents examples of how anonymized network data can both protect privacy and still have utility for security analysis.

In the network security realm, different anonymization techniques provide different levels of privacy protection relative to the organizational security policy in question. The golden rule is that anonymized field values must still be valid field values so processing is invariant to anonymization. Anonymization for network data includes the following techniques: [20]

- <u>Filtering</u> – deletion of field values
- <u>Replacement</u> – pseudo-anonymous permutation mappings or fully-anonymous substitution mapping of field values
- <u>Reduction of Accuracy</u> – approximating data values (examples include truncation or rounding) or mapping of field values into groups
- <u>Adding Noise</u> – adding noise to perturb field values (examples include time shifting)
- <u>Aggregation</u> – summarization of field values with cumulative statistics

## 3. EXPERIMENTAL DESIGN

In this section we describe experiment components and the protocol of how they are used together for producing results for analysis. The components include a packet trace anonymizer tool, a large network traffic dataset, and a standard open-source security analysis tool.

### 3.1 Experiment Protocol

Our intention is to make quantitative measurements across a large dataset so general statements can be made about the effects of anonymization on privacy/analysis tradeoffs when sharing network data. Table 1 is a list of the packet fields and the corresponding 91 experiments we conducted. For each of the 91 experiments, we tested the corresponding anonymization options over all 131 files in the dataset. We developed scripts to automatically configure and execute the anonymization options for each field and then piped the output anonymized packet traces to an intrusion detection system (IDS) to observe alarm results. Processing on each file is an independent replication – no algorithm or parameter linkages exist between each replications.

Since each of the 131 files vary in size, content, and when the packet traces were captured, we developed a uniform way to relate IDS alarm results from different replications. We first established a benchmark number of IDS alarms for each file given no anonymization. Then for each independent replication we measured the deviation from the corresponding file benchmark. For each anonymization option tested, we then statistically processed IDS alarm results relative to the corresponding file benchmark to find the mean, standard deviation, and range.

General intuition is that anonymization is a zero sum tradeoff between privacy and security – the more network data is anonymized for privacy-protection, the less value the network data may be for security analysis. The metric we use as a proxy for security analysis is IDS alarms. We are aware, however, that IDS alarms are not a perfect proxy for security analysis. While less IDS alarms maps to lower levels of security analysis, the relationship of more IDS alarms to security analysis is non-linear. With more IDS alarms, more security analysis may have taken place if (and only if) new information is revealed by the new IDS alarms. However, more IDS alarms may also decrease ability to perform security analysis if the additional alarms are inaccurate or redundant. Despite this additional complexity, IDS alarms do provide a quantitative metric for security analysis and we carefully examine details about the nature of IDS alarms in the experimental results.

**Table 1. Experiments on Single-Fields in Packets Traces**

| Single-Field [layer] | Multi-level Anonymization Option Experiments |
|---|---|
| *(1) Transport Protocol Number [network]* | 12 experiments. Anonymization of (1) all packets; (2) TCP only, (3) UDP only. Multi-level anonymization options: (a) black marker; (b) pure randomization, (c) keyed randomization, (d) bilateral classification [TCP/UDP/ICMP well-known protocols / other]. |
| *(2) Total Packet Length [network & PCAP]* | 4 experiments. Multi-level anonymization options: (a) black marker; (b) pure randomization; (c) keyed randomization; and (d) grouping [0-100, 101-2000, 2001-65536]. |
| *(3) Time-To-Live [network]* | 4 experiments. Multi-level anonymization options: (a) black marker; (b) pure randomization; (3) keyed randomization; (4) grouping [0, 1-32, 33-64, 65-255]. |
| *(4) Type-Of-Service [network]* | 4 experiments. Multi-level anonymization options: (a) black marker (not fragment/fragment), (b) pure randomization, (c) keyed randomization, (d) bilateral classification (0 in bit one sets all bits to 0, 1 in bit one sets all bits to 1). |
| *(5) Fragmentation Flags [network]* | 4 experiments. Multi-level anonymization options: (a) black marker (not fragment/fragment), (b) pure randomization, (c) keyed randomization. |
| *(6) IP Address [network]* | IP address field is anonymized in selected dataset so no anonymization experiments possible. |
| *(7) Ports [transport]* | 30 experiments. Anonymization of (1) both source and destination ports, (2) source ports only, (3) destination ports only. Multi-level anonymization options: (a) black marker [set to 0]; (b) bilateral [below/above 1024]; (c) pure randomization; (d) keyed randomization. |
| *(8) Sequence Number [transport]* | 6 experiments. Multi-level anonymization options: (a) black marker; (b) pure randomization; (c) keyed randomization; (d) grouping [0-1M, 1M-2M, 2M-3M, 3M-max value]. |
| *(9) Window Size [transport]* | 6 experiments. Multi-level anonymization options: (a) black marker; (b) pure randomization; (c) keyed randomization; (d) bilateral classification [below/above 10000]; (e) grouping [0-1024, 1025-8192, 8193-16384, 16385-32768, 32769-65535]. |
| *(10) TCP Flags [transport]* | 11 experiments. Multi-level anonymization options: (a) black marker all flags or black marker each flag individually [URG/ACK/PSH/RST/SYN/FIN], (b) grouping by setting RST/SYN/FIN flags = 0 or by setting URG/ACK/PSH flags = 0, (c) pure randomization, (d) keyed randomization. |
| *(11) Payload [transport]* | Payload field is deleted in selected dataset so no anonymization experiments possible. |
| *(12) Time Stamp [PCAP]* | 10 experiments. Multi-level anonymization options: (a) black marker; (b) time unit annihilation (seconds:microseconds); (c) truncation; (d) enumeration; (e) random time shift; (f) pure randomization; (g) keyed randomization. |

### 3.2 *SCRUB-tcpdump* Tool [28]

We use the *SCRUB-tcpdump* network packet trace anonymization tool due to its flexibility to anonymize all fields and options that provide for different levels of anonymization within each field. As its name suggests, *SCRUB-tcpdump* builds upon the popular *tcpdump* [27] tool for easy data management of packet traces while simultaneously protecting private/sensitive data from being disclosed through the use of anonymization. With *SCRUB-*

*tcpdump*, a user can anonymize fields considered sensitive to multiple desired levels by selecting options that remove all information, add noise, or permute the data.

*SCRUB-tcpdump* is part of a larger effort to build an integrated *SCRUB\** infrastructure for privacy-protected sharing of network data based on consistent multi-level anonymization tools. Other multi-level anonymization tools for network data which use the same algorithms include *CANINE* for NetFlows [4,12] and *SCRUB-PA* for Process Accounting [21,13]. For more details about the anonymization algorithms see [12,13] Three anonymization options are reused multiple times so a general description is provided:

> Black Marker – replacing every value of the field with a predefined constant matching the value type expected in the field. Since all the information in the field is eliminated, this method results in 100% information loss. [12,13]
>
> Pure Randomization – The value in the field is mapped to any valid permutation. To accomplish this, we make use of tables to store mappings between unanonymized and the corresponding anonymized values. Because the creation of the table is dependent upon the state of the pseudorandom number generator and the packet trace being anonymized (specifically field values and the order in which they appear) *mappings will be different every time this algorithm is executed (the same input value will be anonymized to different output values).* If mappings must be consistent between different packet traces that are anonymized we recommend the use of keyed randomization. [12,13]
>
> Keyed Randomization – The mapping from unanonymized to anonymized data is well-defined by a small key. Thus mappings are consistent between different packet traces with this method at different times – as long as the same key is used. The implementation uses keyed hashes (also called keyed-Hash Message Authentication Code or HMACs). The result of HMAC implementation is that data is no longer permuted. However, collisions are low and hence it is nearly a one-to-one correspondence between unanonymized and anonymized values. [12,13]

## 3.3 Network Packet Trace Dataset

The packet trace dataset selected for experimentation is from the Lawrence Berkeley National Laboratory (LBNL), a research institution with a medium-sized enterprise network.[10] Table 2 shows an overview of dataset characteristics. The data was collected over three months (from October 2004 through January 2005) on five separate days. Each day trace covers a range of ten minutes to one hour with 2,000+ hosts monitored. IP is dominant with 96-99% of the packets. The heavy-weight traffic applications are network file systems (NFS and Netware Control Protocol) and systems backup (Dantz and Veritas).[16]

There are important caveats with using this dataset. While this is the only enterprise traffic trace of its magnitude available for open study, it is still only one instance that may or may not generalize to other enterprises. Second, the payload field is deleted and the IP address field anonymized so anonymization experiments on these two fields are not possible. Third, all the data has not yet been released for public study – 11GB of packet header traces in 131 files has been released as of September 2007. Fourth, traffic was captured by tapping links from subnets to main routers so only traffic between subnets is captured.[16] Fifth, sequential scanning activity (both legitimate vulnerability scans and malicious probe scans) create structural relationships that become a de-anonymization threat – for this reason scanning activity is filtered from each packet trace file and instead stored in separate corresponding scan-only files.[17] Sixth, in many files we found truncated UDP packets with length fields we had to recalculated.

**Table 2. Dataset Characteristics** (adapted from [16])

| LBNL PACKET TRACE DATA | DAY 1 | DAY 2 | DAY 3 | DAY 4 | DAY 5 |
|---|---|---|---|---|---|
| DATE | Oct. 4[th] 2004 | Dec 15[th] 2004 | Dec 16[th] 2004 | Jan 6[th] 2005 | Jan 7[th] 2005 |
| DURATION | 10 min | 1 hour | 1 hour | 1 hour | 1 hour |
| PER TAP | 1 | 2 | 1 | 1 | 1-2 |
| SUBNETS | 22 | 22 | 22 | 18 | 18 |
| PACKETS (M) | 17.8 | 64.7 | 28.1 | 21.6 | 27.7 |
| SNAPLEN | 1500 | 68 | 68 | 1500 | 1500 |
| BYTES (GB) | 13.12 | 31.88 | 13.20 | 8.98 | 11.75 |
| CONNECTIONS (M) | 0.16 | 1.17 | 0.54 | 0.75 | 1.15 |
| TRACED HOSTS | 2,531 | 2,102 | 2,088 | 1,561 | 1,558 |
| LBNL HOSTS | 4,767 | 5,761 | 5,210 | 5,234 | 5,698 |
| OTHER HOSTS | 4,342 | 10,478 | 7,138 | 16,404 | 23,267 |
| IP (% packets) | 99% | 97% | 96% | 98% | 96% |
| !IP (% packets) | 1% | 3% | 4% | 2% | 4% |
| !IP% - ARP (% packets) | 10% | 6% | 5% | 27% | 16% |
| !IP% - IPX (% packets) | 80% | 77% | 65% | 57% | 32% |
| !IP% - OTHER (% packets) | 10% | 17% | 29% | 16% | 52% |
| TCP (% bytes/ % connections) | 66% / 26% | 95% / 19% | 90% / 23% | 77% / 10% | 82% / 8% |
| UDP (% bytes/ % connections) | 34% / 68% | 5% / 74% | 10% / 70% | 23% / 85% | 18% / 87% |
| ICMP (% bytes/ % connections) | 0% / 6% | 0% / 6% | 0% / 8% | 0% / 5% | 0% / 5% |

## 3.4 Snort IDS

We selected the Snort IDS [19] for experimentation for two reasons: (1) we want a widely-used IDS so our experiments are reproducible as well as trusted, and (2) we require the ability to examine a standard open-source ruleset in order to understand why certain rules fire during experimentation. The ruleset utilized is the official "Sourcefire VRT Certified Rules (registered user release)" for version 2.24 with every rule turned on.[2] This is the set of rules developed by the Sourcefire company that is continually kept up-to-date to include alerts for the newest and most critical security problems. This ruleset consists of 435 SVT (snort vulnerability team) rules, 98 chain headers, and 16 configuration files.[3]

---

[2] Downloaded July 19, 2007.

## 4. EXPERIMENTAL RESULTS – PRIVACY/ANALYSIS TRADEOFFS FOR EACH FIELD

### 4.1 Transport Protocol Number Field

The network layer Transport Protocol Number Field (8 bits) identifies where the network layer passes/receives information with the transport layer and is important for security analysis because it identifies types of packet flows (connection-oriented TCP flows, connectionless UDP flows, ICMP error messages, etc). We report that snort alarms are closely tied to well-known transport protocols (TCP/UDP/ICMP) such that when these transport protocol numbers are anonymized few alarms trigger except for alarms not related to IPv4 network packets.[4]

Figure 1 shows raw data from the pure randomization anonymization option on the Transport Protocol Number Field. The x-axis is the percent difference in IDS alarms from each of 131 anonymized files as compared to benchmarks for each file without anonymization. This data validates that snort has distinct TCP and UDP alarms that sum to the total number of IDS alarms. When all packets are anonymized, IDS alarms percentage change varies from -100% (0 alarms after anonymization) to +5482% (25,492 more alarms after anonymization). When only TCP packets are anonymized, the percent difference in IDS alarms is negative (between -100% and 0) since TCP traffic dominates UDP traffic and anonymization of TCP packets leaves less UDP packets to trigger alarms. When only UDP packets are anonymized, a minority of files with significant UDP flows have less IDS alarms.

This field is key to snort IDS analysis, every rule in the ruleset we use contains a signature for this field. Table 3 shows that pure randomization of this field creates many snort IDS alarms while removing all information from this field via black marker/keyed randomization/bilateral classification options decreases alarms to zero in the majority of files. Thus anonymization of the Transport Protocol Number Field resembles a zero-sum tradeoff, the more privacy the less security analysis capability and vice versa.

---

[3] Snort ruleset breakdown by configuration files: local.rules=0, bad-traffic.rules=6, scan.rules=12, finger.rules=13, ftp.rules=78, rpc.rules=141, rservices.rules=13, dos.rules=12, ddos.rules=30, dns.rules=21, tftp.rules=15, icmp.rules=22, misc.rules=53, attack.responses.rules=16, other-ids.rules=3, experimental.rules=0.

[4] Examples of data link level alarms that triggered with anonymization of well-known transport protocol numbers include: Point-to-Point Protocol over Ethernet (PPPOE) and Address Resolution Protocol (ARP) alarms. We also had randomization mappings to these transport protocol numbers which triggered alarms: 53 (SWIPE – IP with encryption), 55 (MOBILE – IP mobility), 77 SUN ND protocol, and 103 (PIM – Protocol Independent Multicast).

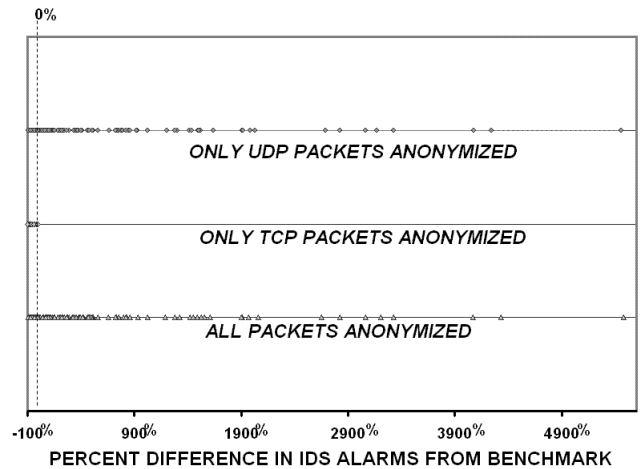

**Figure 1. Transport Protocol Number Field Anonymization: Scatter Plot of Pure Randomization Option**

We also foreshadow about multi-field interactions – preliminary results show that anonymization of the IP transport protocol Number Field has precedence over other packet fields. We find that when an input file with the transport protocol number anonymized via black marker is input to snort, it takes significantly shorter processing time since snort switches processing to signatures exclusively designed for the data link and network layers so deeper transport layer packet inspection is short-circuited.

**Table 3. Effect of Transport Protocol Number Field Anonymization Options**

| ANONYMIZATION OPTION | ALERTS WITH OPTION |
| --- | --- |
| BLACK MARKER | (mean) -93.89% (stdev) 24.04 (range) -100 to 0 |
| PURE RANDOMIZATION | (mean) +549.04% (stdev) 955.77 (range) -89% to +5482% |
| KEYED RANDOMIZATION | (mean) -93.78% (stdev) 24.04 (range) -100% to 0% |
| BILATERAL CLASSIFICATION | (mean) -93.89% (stdev) 24.04 (range) -100% to 0% |

### 4.2 Total Packet Length Field

The network layer[5] Packet Length Field (16 bits) denotes the length of the IP datagram including all headers/data. Packet length is important for security analysis because some security events have uniquely sized packets (Slammer Worm at 404 bytes, Nachi Worm at 92 bytes, video slicing applications that create files at exact MB sizes). Packet lengths can also be combined with other information to indirectly infer information like the identity of a file/server/client; or a protocol event (e.g. HTTP item size).

---

[5] PCAP also provides "packet length" and "captured packet length" fields which we did not use – the "captured packet length field" for the selected dataset was recalculated since the payload field was deleted.

We report from 4 experiments over all 131 files that anonymization of the Total Length Field has an intermediate effect on snort IDS alarms. Snort rules check length value validity in order to prevent buffer overflow attacks and these alarms may be triggered with anonymization options. For instance, total IP datagram length cannot be less than the IP header length:

```
[116:3:1] (snort decoder) WARNING: IP dgm len < IP hdr len!
```

Table 4 shows that eliminating information from this field via the black marker anonymization option greatly increases snort alarms due to added uncertainty and this decreases security analysis capability. The grouping option also adds uncertainty/noise to field values which significantly increases snort alarms. Pure and Keyed Randomization options only slightly increase the number of snort alarms. Thus, anonymization of the Packet Length Field resembles a zero-sum tradeoff, the more privacy the less security analysis capability and vice versa.

**Table 4. Effect of Packet Length Field Anonymization Options**

| ANONYMIZATION OPTION | ALERTS WITH OPTION |
|---|---|
| BLACK MARKER | (mean) +32733.23% (stdev) 49102.97 (range) 0 to +286205 |
| PURE RANDOMIZATION | (mean) +20.44% (stdev) 33.50 (range) 0% to +206% |
| KEYED RANDOMIZATION | (mean) +4.69% (stdev) 23.43 (range) 0% to +240% |
| GROUPING | (mean) +250.82% (stdev) 896.92 (range) 0% to +9637% |

## 4.3 Time-To-Live Field

The network layer Time-To-Live (TTL) Field (8 bits) denotes the limited lifetime of a packet in a network. TTL is important for security analysis since: (1) popular operating systems can be fingerprinted (identified) from well-known initial TTL values;[6] (2) path hop lengths can be estimated by subtracting observed TTL values from the closest initial TTL value; (3) route changes can be detected from TTL changes; (4) TTL can be used as a covert channel; and (5) TTL has been proposed for IP traceback.[30]

We report from 4 experiments over all 131 files that anonymization of the TTL Field has no effect on snort IDS analysis. There are no TTL signatures in the snort v.2.24 ruleset, however, two rules specific to TTL of DNS packets at the application layer (not IPv4 header). Thus, the TTL Field (as a single field) may be anonymized to protect privacy without any impact on security analysis using a snort IDS.

## 4.4 Type-of-Service Field

The network layer Type-of-Service (TOS) Field (8 bits) is used to specify the treatment of an IP datagram during its transmission through the Internet. The first 3 bits allow a network administrator to assign precedence values from 0 (default) to 7 to classify and prioritize types of traffic. Bits 3/4/5 represent requests for low delay (D), high throughput (T), high reliability (R) respectively. TOS is important for security analysis for the following reasons:

(1) it reveals whether an application uses TOS (which is not typically implemented), (2) differences in ICMP error messages TOS Field may identify/fingerprint types of routers[7], and (3) as a user-defined field reveals user behavior. Thus, the TOS Field (as a single field) may be anonymized to protect privacy without any impact on security analysis using a snort IDS.

## 4.5 Fragmentation Flags Field

The network layer Fragmentation Flags Field (3 bits) is used with the Fragmentation Offset Field to provide the ability to transmit and reassemble datagram fragments. The "Fragment Flag/Bit" signals routers not to fragment the current datagram. If this cannot be done an error message is returned. The "More Fragments Flag/Bit" is used to signal whether the datagram being sent is actually a fragment of a larger datagram or not fragmented (first and only datagram). This field is important for security analysis because fragmentation can be utilized to circumvent filtering rules.[8]

We report from 4 experiments over all 131 files that anonymization of the Fragmentation Flag Field has minimal effect on snort alarms. A small number of alarms are triggered by fragmented datagrams packets using the UDP (signature for filtering circumvention using fragmentation) and IGMP transport protocols (signature for IGMP denial-of-service attack)[9], both multi-field effects. Thus, the Fragmentation Flags Field may be anonymized to protect privacy with minimal impact on security analysis using a snort IDS.

## 4.6 Ports Field

The transport layer Ports Field (16 bits source/destination) identifies a specific process socket where network messages are to be forwarded. Ports are important for security analysis because there is an underlying *port number assumption* in the ability to map port numbers to network services. Thus packets using certain ports are assumed to be running the corresponding service. If coordinated between hosts in advance, services can run with any port numbers, change dynamically, or be tunneled within traffic over unrelated ports. Examples include covert port knocking, backdoors installed by attackers on nonstandard ports to facilitate return/control, and P2P traffic tunneled within HTTP. However, the reality is that most malicious traffic follows the port number assumption making port numbers arguably the most important information used by security analysts.

---

[6] Over fifty examples of common initial TTL values:
<http://members.cox.net/~ndav1/self_published/TTL_values.html>

[7] <http://seclists.org/nmap-hackers/2000/0332.html>

[8] One example is to set the value of the Fragment Offset on a second packet so low that instead of appending the second packet to the first packet, it actually overwrites the data and part of the TCP header of the first packet. A packet filter will see that the Fragment Offset is greater than zero on the second packet, deduce that the second packet is a fragment of another packet, and thus not check the second packet against the rule set. Examples of other fragmentation attacks can be found in [32].

[9] A fragmented IGMP packet may allow the TCP/IP stack to gain access to invalid segments of computer memory.
<http://support.microsoft.com/default.aspx?scid=kb;en-us;238329&sd=tech>

We report that for our test environment port field anonymization has no effect on the number of snort IDS alarms. This is an example where port anonymization can protect privacy and yet have no effect on IDS security analysis. Examining the snort ruleset we find that the snort signatures which include ports are triggered by both port number and payload data. Since the dataset we use has stripped the payload field, there could be no alarm matches between port numbers and payload data with or without anonymization. Future experiments are planned with datasets that include the payload field.

We also report the effect of Snort preprocessor plug-ins for performing behavior analysis on network flows. In our default configuration, the "stream4" preprocessor is loaded. The preprocessor configuration allows users to specify ports where legitimate servers are operating (e.g., web servers on port 80) in which case the preprocessor will switch to identifying suspicious flow behavior on these specific ports instead of independently processing individual packets. The Snort preprocessor ability may be useful with anonymized shared data if the ports field is either unanonymized or if server ports are otherwise identifiable.

## 4.7 Sequence Number Field

The transport layer Sequence Number Field (32 bits) is used to acknowledge receipt of data which also maintains sequential order. Sequence numbers are important for security analysis because: (1) the PRNG used to generate sequence numbers cannot be easily modified and since different systems have different generating functions this provides reliable operating system fingerprinting [29], and (2) there are numerous attacks based on sequence numbers.[10]

We report from 6 experiments over all 131 files that anonymization of the sequence numbers has no effect on snort IDS analysis. Eight rules in previous snort rulesets used sequence numbers in their signatures but none of these rules are contained in the v.2.24 snort ruleset. Thus, the Sequence Number Field (as a single field) may be anonymized to protect privacy without any impact on security analysis using a snort IDS.

## 4.8 Window Size Field

The transport layer Window Size Field (16 bits) provides the upper limit of how many bytes a sender can transmit before receiving an acknowledgment. Window size is important for security analysis because different operating systems can be fingerprinted due to default and upper limit window size values.[11]

We report from 6 experiments over all 131 files that anonymization of the Window Size Field has no effect on snort IDS analysis. One rule in a previous snort ruleset used window size in its signature but this rule is not contained in the v.2.24 snort ruleset. Thus, the Window Size Field (as a single field) may be anonymized to protect privacy without any impact on security analysis using a snort IDS.

## 4.9 TCP Flags Field

The transport layer TCP Flags Field (8 bits) consists of 8 flag/control bits. For this paper we consider the six flags (FIN, SYN, RST, PSH, ACK, URG) related to TCP connection dynamics. TCP flags are important for security analysis because they make it possible to identify certain types of denial-of-service and session hijack attacks. Due to ambiguities in TCP implementations for different operating systems, TCP flags can be used to fingerprint different operating systems.[12] Probe packets with unusual TCP flag combinations have been known to circumvent packet filters.[15]

We report from 11 experiments over all 131 files that anonymization of the TCP Flags Field had little or no effect on snort IDS analysis. There are three types of rules in the v.2.24 snort ruleset which use TCP flags in their signature – alarms for (1) DOS BGP spoofed connection reset attempt, (2) miscellaneous source ports (20, 53), and (3) scan probes. We found these rules did not fire due to random mappings.

## 4.10 Time Stamp Field

PCAP time stamps packets (seconds:microseconds) as captured providing insight into traffic dynamics such as interarrival times, unambiguous matching of packets with acknowledgments, and detecting packet duplication and reordering.[17] Time stamps are important for security analysis because: (1) they facilitate security event correlation, and (2) anomaly-based network IDSs create profiles of normal traffic behavior over time and trigger alarms when unusual traffic patterns occur. Time stamps can also be used to fingerprint computers based on clock drift[8] and to indirectly infer information when combined with other data.

We report from 10 experiments over all 131 files that time stamp anonymization has no effect on snort IDS analysis. There are no snort v.2.24 rules containing timestamps. Thus, time stamps (as a single field) may be anonymized to protect privacy without any impact on security analysis using a snort IDS. Anomaly-based IDSs which attempt to correlate events in time to find unusual (potentially malicious) activity may have different results.

## 5. CONCLUSIONS

Trust management is an active research area. This work is an initial step toward facilitating trusted sharing of network packet traces for distributed security analysis based on the use of anonymization. While anonymization tradeoffs between privacy protection versus security analysis have been speculated upon by many researchers, this is the first work to empirically characterize these tradeoffs for sharing network data — specifically single-fields within packet traces. Future work will study tradeoffs from anonymizing multiple-fields within packet traces.

Our results validate intuition that in some cases anonymization increases privacy protection at the expense of security analysis capability. For instance, protecting privacy with anonymization options on the Transport Protocol Number and Packet Length Fields come at the expense of security analysis capability. For instance, protecting privacy with anonymization options on the Transport Protocol Number and Packet Length Fields come at the expense of security analysis capability (increased uncertainty which significantly increases the number of triggering snort alarms). However, the overall privacy/analysis tradeoff is complex and not always a zero sum tradeoff. We found applying anonymization options on eight other single-fields within a packet can simultaneously protect privacy and support effective security analysis – anonymization on these eight single-fields does not affect snort IDS alarms.

---

[10] For example, the Sequence Number Prediction attack is based on the initial sequence numbers (ISN) used by TCP, which should be random. However, Berkeley Unix starts with an ISN of 1 and increments it a fixed number of times per second and per connection. It is therefore possible to estimate the next ISN that will be used by connecting to the server, recording the ISN and then measuring the time to the next connection.

[11] Twenty five examples of common window size values: <http://www.honeynet.org/papers/finger/traces.txt>

[12] Four examples of OS fingerprinting using TCP flags: <http://www.gray-world.net/papers/ambiguitiesintcpip.txt>